\documentclass[arps,prd,twocolumn,twoside,floatfix,noshowpacs,nofootinbib]{revtex4}
\pdfoutput=1

\usepackage{xcolor,amsmath}
\usepackage{bm}
\usepackage{ulem}

\newcommand{\ba}{\begin{eqnarray}}
\newcommand{\ea}{\end{eqnarray}}
\newcommand{\be}{\begin{equation}}
\newcommand{\ee}{\end{equation}}

\newcommand{\jcap}{{J.~Cosm.~Astrop.~Phys.}}

\newcommand{\aap}{{Astron.~Astrophys.}}
\newcommand{\apjl}{{Astrophys.~J.~Lett.}}
\newcommand{\apjs}{{Astrophys.~J.~Supp.}}
\newcommand{\aj}{{Astron.~J.}}
\newcommand{\mnras}{{Mon.~Not.~R.~Astron.~Soc.}}

\usepackage{graphicx}
\usepackage[large]{subfigure}
\usepackage{amssymb, amsmath}
\usepackage[amssymb]{SIunits}
\usepackage{epstopdf}
\usepackage{hyperref}
\usepackage{url}
\usepackage{aas_macros}
\usepackage{natbib}

\begin{document}
\title{Can Conformal and Disformal Couplings Between Dark Sectors Explain the EDGES 21cm Anomaly?}
\author{Linfeng Xiao\footnote{hartley@sjtu.edu.cn}}
\author{Rui An}
\author{Le Zhang\footnote{lezhang@sjtu.edu.cn}}
\affiliation{IFSA Collaborative Innovation Center, School of Physics and Astronomy, Shanghai Jiao Tong University, Shanghai 200240, China}
\author{Bin Yue}
\author{Yidong Xu}
\affiliation{Key Laboratory for Computational Astrophysics, National Astronomical Observatories, Chinese Academy of Sciences, Beijing 100101, China}
\author{Bin Wang\footnote{wangb@yzu.edu.cn}}
\affiliation{Center for Gravitation and Cosmology, YangZhou University, Yangzhou 225009, China}
\begin{abstract}
The recently announced result by EDGES points an unexpected excess in the 21 cm global brightness temperature from cosmic dawn at $z\sim 17$, potentially indicating new phenomena beyond the $\Lambda$CDM model. A generic cosmological model which allows conformal and disformal couplings between dark matter and dark energy is employed to investigate the impact on the 21cm absorption signal and understand the EDGES anomaly.  After exploring a wide range of parameter space for couplings, we find that the coupling effects can lead to a moderate change in the Hubble parameter while a negligible change in the spin temperature in the early Universe. Consequently, the decrease of the Hubble parameter from the mixed conformal and disformal couplings can reproduce the 21cm absorption approximately in consistent with the EDGES result at $z=17.5$. However, there is still tension in corresponding parameter space between EDGES and other cosmological observations for this model.
\end{abstract}

\maketitle

\section{Introduction}
\label{sec:intro}
Recently, the Experiment to Detect the Global Epoch of Reionization Signature (EDGES) reported the first detection of 21 cm hydrogen absorption signal relative to the Cosmic Microwave Background (CMB) centered at $z\sim17$~\cite{EDGES}, opening a new window to the early universe. However, the amplitude of the EDGES signal is more than a factor of two greater than the largest theoretical predictions~\cite{Barkana2018Nature}. If confirmed, this anomalously strong 21 cm absorption will indicate some new underlying physics at work.

In the so-called concordance $\Lambda$CDM cosmology, neutral hydrogen (HI) gas temperature evolves as $(1+z)^2$ due to adiabatic expansion after baryon decoupling from CMB photons at $z\sim150$, while the CMB temperature evolves steadily as $\propto (1+z)$. After this decoupling, the decreasing density of the gas gradually becomes insufficient to maintain the collisional coupling between the hydrogen spin temperature and the gas kinetic temperature, and therefore the spin temperature eventually re-equilibrates with the CMB temperature at $z\lesssim 40$. Until first star formation (most likely at $z\lesssim 30$), the Ly$\alpha$ photons emitted from these luminous sources coupled the 21 cm spin temperature to the gas kinetic temperature through Wouthuysen-Field effect~\cite{Wouthuysen1952,Field1959}, making an absorption feature in 21 cm brightness temperature before the intergalactic medium was significantly heated. Finally, the increasing X-ray emissions from these first stars would heat the gas above the CMB, leading to a 21 cm emission signal.

Thus, the EDGES-detected absorption feature of 21 cm signal at the redshifts of $15\lesssim z\lesssim20$ is consistent with the prediction in the $\Lambda$CDM model~\cite{Cohen2017}, whereas the best-fitted amplitude of the absorption ($T_{21}=-500_{-500}^{+200}$ mK at 99\% confidence level) far exceeds the expectation, with about a factor of two (corresponding to $3.8\sigma$) greater than the largest prediction of $T_{21} \simeq-209$ mK~\cite{Barkana2018Nature} in the standard model.

Generally, the brightness temperature of the observed 21 cm signal is related to the Hubble parameter $H(z)$, and the ratio between the temperature of the background radiation $T_\gamma$ and the spin temperature of the gas $T_{\rm s}$, i.e.,
\be\label{eq:t21-1}
T_{21} \propto \frac{1}{H(z)}\left(1- \frac{T_\gamma (z)}{T_{\rm s}(z)}\right)\,,
\ee
According to Eq.~\ref{eq:t21-1}, many mechanisms recently have been proposed to explain this significant anomaly, ~\cite{Barkana2018Nature,ML2018,FBC2018,BHKM2018,BORV2018,FHH2018,LC2018,CDGMS2018,ECL2018,Costa2018,HB2018,CKNT2018}, such as (a) reducing the spin temperature $T_{\rm s}$, or (b) increasing the effective background radiation temperature $T_\gamma$ such that $T_\gamma > T_{\rm CMB}$, or (c) reducing the Hubble parameter.

In the first attempt, one  introduced new baryon-dark matter (DM) interactions which can transfer the baryonic kinetic energy into DM so as to cool down the gas efficiently. However, this attempt is constrained by cosmological observations severely~\cite{Barkana2018Nature,ML2018,FBC2018,BHKM2018,BORV2018,FHH2018,LC2018, CKNT2018}, one has to ``fine-tune'' the properties (e.g., mass and cross-section) of the most prevailing DM candidates -- weakly interacting massive particles (WIMPs). Ref.~\cite{Barkana2018Nature} sets up an upper limit on DM particle mass in light of EDGES, where the allowed mass (less than a few GeV) is much lighter than that expected for WIMPs.

Alternatively, it is possible to raise the radiation temperature by adding extra radio backgrounds in the early universe. Some examples for this attempts were given in~\cite{CDGMS2018,ECL2018}, where extra photons are expected to be produced either by accretion of intermediate mass black holes or by light WIMPs annihilation or decay.

Besides these two scenarios, as $T_{21}$ linearly depends on the value of $1/H(z)$ as seen in Eq.~\ref{eq:t21-1}, the EDGES anomaly thus can be, of course, due to a smaller value of $H(z)$. The first attempt to modify $H(z)$ and  explain the anomaly was proposed by introducing a phenomenological interaction between dark sectors \cite{Costa2018}. In this study the dark energy model was not specified and its equation of state was treated as a free constant smaller than $-1/3$. Recent discussions on the interaction between dark sectors can be found in the review \cite{WAAP2016} and references therein. Further attempt in this direction was also proposed by considering the early dark energy \cite{HB2018} with the phenomenological  model \cite{KK2016}. All these models are qualitatively shown effective in altering $H(z)$, reconciling the tension between $\Lambda$CDM predictions and the EDGES result. However, it is still unclear that whether the parameters adopted in these models to fit EDGES are consistent with other recent cosmological measurements such as the CMB data from Planck satellite~\cite{Planck2015-13} etc.

In a recent paper, it was argued that the tension between the EDGES signal and the $\Lambda$CDM prediction can be released by introducing an interaction between dark matter and vacuum energy \cite{WZ2018}. This study is in the spirit of the interaction between dark sectors, but assuming the vacuum energy as the candidate for dark energy. It was reported that the EDGES measurement can marginally improve the constraint on parameters that quantify the interacting vacuum, while the combined dataset favors the $\Lambda$CDM at 68\% CL. Simply considering the vacuum energy as dark energy suffers serious theoretical problems. This is the reason that a dynamical scalar field is called for. Recently a new  perspective on understanding the vacuum energy problem  in the expanding universe with scalar field was suggested in \cite{HQ2016}.

Considering dark matter and dark energy occupying 95\% of our universe, studying the interaction between them is an intriguing and rational step towards a more comprehensive understanding of the nature. In this paper we will focus on a generic interacting model between dark sectors. We assume the dark energy as a canonical quintessence scalar field to avoid the vacuum energy problem and being conformally and disformally coupling with dark matter. Conformally coupled dark energy models in the Einstein frame  have the relation to the modified gravity in the Jordan frame, which have been exhaustively explored, see  \cite{WAAP2016} and references therein.  Recently, interacting dark energy models with a disformal coupling have been recently confronted with astronomical observations \cite{BM2015,MB2017,BM2017}. In this work, we will re-examine this generic coupled dark energy model and investigate its impact on  21 cm absorption signal and give the proof-of-the-concept explanation on the EDGES anomaly.

The paper is organized as follows. In Sec.~\ref{sec:physics}, we give a brief introduction to the generic DM-DE coupling model and the 21 cm brightness temperature. In Sec.~\ref{sec:results} we consider the conformal coupling, disformal coupling and the combined conformal and disformal couplings on the influence of the 21 cm brightness temperature.  Finally, we give conclusions and discussions in Sec.~\ref{sec:conclusions}.

\section{DM-DE coupling Model and 21 cm brightness temperature}
\label{sec:physics}

In this section we first briefly review the generic model on the interaction between dark matter and dark energy, which has been thoroughly discussed in \cite{BM2015,MB2017,BM2017}. In the Einstein frame, the scalar-tensor theory action of the model reads
\be
\label{action}
\begin{split}
	\mathcal{S} =& \int d^4 x \sqrt{-g} \left[ \frac{M_{\rm Pl}^2}{2} R - \frac{1}{2} g^{\mu\nu}\partial_\mu \phi\, \partial_\nu \phi - V(\phi) + \mathcal{L}_{\rm SM}\right]\\
	&+ \int d^4 x \sqrt{-\tilde{g}} \mathcal{\tilde{L}}_{\rm DM}\left(\tilde g_{\mu\nu}, \psi\right)\,,
\end{split}
\ee
where the reduced Planck mass is $M_\text{Pl}=2.4\times 10^{18}$ GeV. The Lagrangian $\mathcal{L}_{\rm SM}$ represents the visible sector from the standard model (SM), and dark energy is described by a quintessence scalar field $\phi$ with a potential $V(\phi)$. The last term ($\mathcal{\tilde{L}}_{DM}$) in the action corresponds to the dark matter sector which depends on the metric with the form
\be
\label{DM metric}
\tilde g_{\mu\nu} = C(\phi) g_{\mu\nu} + D(\phi)\, \partial_\mu\phi\, \partial_\nu \phi\,,
\ee
where $C(\phi)$ and $D(\phi)$ denote the conformal and disformal coupling factors, respectively. We see now that dark matter particles follow geodesics determined by $\tilde g_{\mu\nu}$ and that various aspects of these particles, for instance their mass, will depend on the dark energy field.
A recent quantum understanding on the interaction between dark matter and dark energy argued that there is no significant interaction between heavy dark matter and light dark energy at the microscopic level from the normal perturbative quantum field theory. Interaction between dark sectors can be consistent with quantum field theory if dark energy and a fraction of dark matter are very light axions \cite{DAHK2016}. Here dark matter particles following dark energy field satisfies the microscopic level requirement.

In general, the interaction between DE and DM can be described by \cite{BM2017},
\be
\label{coupling function}
Q=\frac{C_{,\phi}}{2C}T_{\rm DM}+\frac{D_{,\phi}}{2C}T_{\rm DM}^{\mu\nu}\nabla_\mu\phi\nabla_\nu\phi-\nabla_\mu\left[\frac{D}{C}T^{\mu\nu}_{\rm DM}\nabla_\nu\phi\right],\,
\ee
where the subscript \{$,\phi$\} denotes the derivative with respect to $\phi$. The pressureless DM has $T_{\rm DM}^{\mu\nu}$ to be its perfect fluid energy momentum tensor, and $T_{\rm DM}$ is the corresponding trace.

On the other hand, DE is described by a scalar field $\phi$ which obeys the modified Klein-Gordon equation and is now coupled to DM via the coupling function $Q$. Assuming a flat universe under the standard Friedmann-Robertson-Walker (FRW) metric with the line element $ds^2 = g_{\mu\nu}dx^{\mu} dx^{\nu} = a^2(\tau)\left(-d\tau^2 + \delta_{ij} dx^i dx^j\right)$, one has
\be
\label{KG-equation}
\phi^{\prime\prime} + 2 \mathcal{H} \phi^{\prime} + a^2 V_{,\phi} = a^2 Q\;,
\ee
where $a$ is the scale factor of the universe, and the prime ($'$) is the derivative with respect to the conformal time. The conformal Hubble parameter here is $\mathcal{H}=a^\prime/a$, which is related to the Hubble parameter by $H(z)=\mathcal{H}/a$. The non-conservation of the DE must be mirrored in the DM species ``$c$'',
\be
\label{conservation_matter}
\rho_c^\prime + 3\mathcal{H}\rho_c = -Q\phi^{\prime}\,.
\ee

The background form of $Q$ in the FRW universe can be simplified as
\be
\label{Q}
Q=\frac{2D\left(\frac{C_{,\phi}}{C}{\phi^\prime}^2 + a^2V_{,\phi} + 3\mathcal{H}\phi^\prime\right) - a^2C_{,\phi} - D_{,\phi}{\phi^\prime}^2}{2\left[a^2C + D\left(a^2\rho_c - {\phi^\prime}^2\right)\right]} \rho_c .
\ee

The nature of the energy transfer process will be determined on the forms of $C(\phi), D(\phi), V(\phi)$. To be consistent with the discussions in \cite{BM2015,MB2017,BM2017}, we choose exponential functional forms as
\be
\label{coupling_choice}
C(\phi)=e^{2\alpha\kappa\phi},\;D(\phi)=D_m^4 e^{2\beta\kappa\phi},\;V(\phi)=V_0^4 e^{-\lambda\kappa\phi},
\ee
where $\alpha,\,D_m,\,\beta,\,V_0,$ and $\lambda$ are constants and $\kappa\equiv M_\text{Pl}^{-1}$.

For standard matters, such as the relativistic species (``$r$'') and the baryons (``$b$''), we  assume them uncoupled with the scalar field, and hence the evolutions of their energy densities read
\ba\label{BR}\nonumber
\rho_b^\prime + 3\mathcal{H}\rho_b &=& 0\,, \\
\rho_r^\prime + 4\mathcal{H}\rho_r &=& 0\,,
\ea
Finally, with these ingredients, the Friedmann equation takes the standard form:
\be
\label{Friedmann Equation}
\mathcal{H}^2 = \frac{\kappa^2}{3}a^2\left(\rho_\phi + \rho_b + \rho_r + \rho_c \right)\,,
\ee
with $\rho_\phi={\phi^\prime}^2/2a^2+V(\phi)$.

Next, we will investigate the impact of the generic interacting model on 21 cm brightness temperature by considering the conformal coupling, disformal coupling and the mixed conformal and disformal couplings. Since the EDGES only measures the sky-averaged global 21 cm signal rather than the fluctuations, we neglect effects from perturbations~\cite{BM2015} and concentrate only on the background evolution.

Compared with the standard $\Lambda$CDM model, it is obvious that the interaction between dark sectors modify the  expansion history and in turn the 21cm brightness temperature. This was disclosed in \cite{Costa2018} and \cite{WZ2018}. However how the interaction influence the temperatures deviation from that of the $\Lambda$CDM model to leave imprints on the gas temperature $T_{\rm g}$, spin temperature $T_{\rm s}$ and, consequently, 21cm brightness temperature $T_{21}$ is not clear to us.


The evolution history of the universe can be obtained by  solving Eqs.~\ref{KG-equation},~\ref{conservation_matter},~\ref{Q},~\ref{coupling_choice},~\ref{BR} and~\ref{Friedmann Equation}  using the modified  CAMB code~\cite{LCL2000}. The evolutions of the ionization fraction  and the gas temperature are similar to the $\Lambda$CDM model, but the Hubble parameter therein is corrected. The evolution of the gas temperature is governed by adiabatic and Compoton cooling processes~\cite{Seager1999}, if the X-ray heating is negligible, which reads
\be
\label{Tgas Equation}
\frac{dT_{\rm g}}{dz} = \frac{8\sigma_{\rm T}a_{\rm R}
	T_{\rm \gamma}^4}{3H(z)(1+z)m_{\rm e}c}\,
  \frac{x_{\rm e}}{1+f_{\rm He}+x_{\rm e}}\,(T_{\rm g} - T_{\rm \gamma})
+ \frac{2T_{\rm g}}{(1+z)}\,,
\ee
where $\sigma_T$ is the Thomson scattering cross-section, $a_R$ the radiation constant, $m_e$ the electron mass, $c$ the speed of light, $x_e(z)$ ionization fraction of hydrogen, $f_{\rm He}(z)$ the fraction of helium abundance, and the radiation temperature in our scenario is expected to be the same as the CMB temperature $T_{\gamma} (z)  = T_{\rm CMB}=2.7255 (1+z) \, {\rm K}$.

To calculate the 21 cm absorption signal, let us begin with the radiative transfer equation in the Rayleigh-Jeans limit. Assuming homogeneous HI spin temperature, the brightness temperature of the observed radiation field reads~\cite{Zaldarriaga2004}:
\be
T_b(z,\nu)  =  T_{\rm CMB}(z)e^{-\tau_\nu} + T_{\rm s}(z)(1-e^{-\tau_\nu})\label{Tb}\,
\ee
where $\nu$ denotes the observed frequency and $\tau_\nu$ is the optical depth of inter-galactic medium at frequency $\nu$. $T_{\rm s}(z)$ and $T_{\rm CMB} (z)$  stand for the spin temperature and CMB temperature at the redshift $z$, respectively. Therefore, the 21 cm brightness temperature relative to the CMB temperature is
\be
T_{21}(z) \approx \frac{T_{\rm s}(z) - T_{\rm CMB}(z)}{1+z}\tau_{\nu_0}(z)\,, \label{T21 Equation}
\ee
where
\be
\tau_{\nu_0}  =  \frac{3c^3\hbar A_{10}n_{HI}}{16k_B\nu_0^2T_{\rm s}(z)H(z)}.  \label{tau Equation}
\ee
$\nu_0 = 1420.4$ MHz is the rest-frame hyperfine transition frequency, $\hbar$ the reduced Planck constant, $k_B$ the Boltzmann constant and $n_{\rm HI}$ the number density of background neutral hydrogen. $A_{10} = 2.87\times 10^{-15}s^{-1}$ is the spontaneous emission coefficient from the triplet to the singlet state of neutral hydrogen atoms. Here we have assumed homogeneous gas distribution with uniform spin temperature, but note that the gas density fluctuates, and the adiabatic heating of over-dense gas would result in a decreased optical depth and hence a lower level of the 21 cm global signal~\cite{XYC2018}.

The spin temperature is determined by three effects, including the radiative coupling to the CMB, the Wouthuysen-Field and collisional coupling to the gas kinetic temperature $T_{\rm g}$. It can be well-approximated by~\cite{Zaldarriaga2004}
\ba
T_{\rm s} & = & \frac{T_{\rm CMB}+y_{\rm c} T_{\rm g}+y_{\rm{Ly\alpha}}T_{\rm{Ly\alpha}}}{1+y_{\rm c}+y_{\rm{Ly\alpha}}}\,, \label{Ts Equation}\\
y_{\rm c} & = & \frac{C_{10}}{A_{10}} \frac{T_\star}{T_{\rm g}}\,, \label{y_c Equation}\\
y_{\rm{Ly\alpha}} & = & \frac{P_{10}}{A_{10}} \frac{T_\star}{T_{\rm{Ly\alpha}}}\,, \label{y_Lya Equation}
\ea
where $y_{\rm c}$ is the collisional coupling coefficient, $C_{10}$ the collisional de-excitation rate of the triplet hyperfine level, and the energy of 21cm photons is $T_\star=h\nu_0/k_{\rm B}=0.068$K. $y_{{\rm Ly\alpha}}$ is the Ly$\alpha$  coupling coefficient, $P_{10}\approx 1.3\times 10^{-12}J_{-21}s^{-1}$ is the indirect de-excitation rate due to the absorption of Ly$\alpha$ photons.

Here, $J_{-21}$ denotes the Ly$\alpha$ background in units of $10^{-21}$ erg s$^{-1}$cm$^{-2}$Hz$^{-1}$sr$^{-1}$. It is determined from the global star formation history before reionization completed (see details in~\cite{CM2003}), including first stars and first galaxies. Unfortunately, due to the lack of observations of galaxies at $z\gtrsim10$, galaxy properties like the star formation efficiency, the stellar initial mass function and the stellar metallicity are all poorly known, there are large uncertainties on the Ly$\alpha$ background. The values of $J_{-21}(z)$ at redshifts $z<20$ are estimated by using the simulations~\cite{CM2003}, and can be simply extrapolated to high redshifts. We also set  $T_{{\rm Ly}\alpha}=T_{\rm g}$, which is a good approximation for the period of interest (z$\sim$20)~\cite{FOB2006}. We have checked that the results produce a consistent reionization history with recent Planck measurements~\cite{Planck2015-13} in the $\Lambda$CDM frame, computed with the modified Recfast code~\cite{CT2011,Seager1999,Rubino-Martin2010,Chluba2010,CVD2010}.

\section{Results}\label{sec:results}
In this section, we present the detailed analysis of the DM-DE coupling model and investigate its impacts on the 21 cm absorption signal. The joint analysis~\cite{Planck2015-13} of the latest high precision observations from the CMB and other low redshift data gives a flat universe with $\Omega_{\rm c}=0.27$, $\Omega_\Lambda (i.e., \Omega_\phi)=0.68$, $T_{\rm CMB}=2.7255$K and $h=0.6727$ as well as $\Omega_{\rm b}\simeq1-\Omega_\Lambda-\Omega_{\rm c}$. The evolutions of baryon and relativistic species are fixed same as the ones in $\Lambda$CDM ($\rho_b \propto a^{-3}$ and $\rho_r \propto a^{-4}$) since the DE-DM interaction has no effects on baryons and photons.

For evaluating the background evolution for $\rho_{c}$, $\rho_{\phi}$, we have to choose the initial conditions for $\rho_{c}$, $\rho_{\phi}$ and $\phi'$. We first use a {\it shooting algorithm}~\cite{COE2015} which has been implemented in CAMB in order to find appropriate initial conditions for $\rho_{c}$, $\rho_{\phi}$ such that the predicted values of such two parameters at present day are given by the observed ones. Furthermore, the initial value of $\phi'$ has to be also specificed. We choose $\phi'=0$ at $a=10^{-9}$ in this study, which is essentially because that $\rho_\phi \propto \phi'^2/2a^2$ and we assume $\rho_\phi$ at the very early universe is essentially negligibly small. More explicitly, we have verified that, the predicted 21 cm signal is rather stable over a broad range of $\phi'$ ($0\leq\kappa\phi'\lesssim10^{-5}\rm Mpc^{-1}$) for the {\it uncoupled} and {\it conformal} couplings. However, if we increase the initial value of $\kappa\phi'$ from zero by a tiny amount in {\it disformal} and {\it mixed} couplings, e.g.  $\kappa\phi'\geq10^{-14}\rm Mpc^{-1}$ (i.e., $\kappa\dot{\phi}\geq10^{-5}\rm Mpc^{-1}$), the resulting $\rho_c$ and $\rho_\phi$ after recombination will be significantly disfavored by the current observations at above 3-$\sigma$ level. As such, for simplicity, we fix $\phi'=0$ at $a=10^{-9}$ for all the coupling models to solve for the background evolution.

To be consistent with the units chosen in CAMB where $\kappa$ is absorbed into the energy density $\rho$ and $H(z)$ is in units of $\rm Mpc^{-1}$, $\rho$ is thus in the same units of $H^2$. We also know that $D$ is in units of $\rho_c^{-1}$ and $D\propto D_m^4$ in terms of Eqs.~\ref{Q} \&~\ref{coupling_choice}, $D_m$ is thus expressed in units of ${\rm Mpc}^{1/2}$, which is different from the units convention in~\cite{BM2017}. ${V_0}^{4}$ in Eq.~\ref{coupling_choice} is adopted to ${V_0}^{4} = \rho_{\Lambda}(z=0)$. Note that, setting all the coupling parameters ($\alpha,\beta, D_m, \lambda$) to be $0$ means a return to the standard $\Lambda$CDM model.

\begin{table}
	\begin{center}
		\begin{tabular}{ c|  c  c c c}
			\hline
			\hline
			Parameter& \it{uncoupled}  &\it{conformal} & \it{disformal} & \it{mixed} \\
			\hline
			$\lambda$ & 1.3 & 1.3 & 1.3 & 1.6 \\
			$\alpha$  & 0 & -0.1 & 0 & -0.2 \\
			$\beta$ & 0 & 0 & 0 & 0.2  \\
			$D_m/{\rm Mpc}^{1/2}$ & 0 & 0 & 75 & 75 \\
			\hline
			\hline
		\end{tabular}
	\end{center}
	\caption{\label{table:fid} The fiducial parameter setting for {\it uncoupled, conformal, disformal} and {\it mixed} cases.}
\end{table}

\subsection{Impacts of DM-DE coupling on $H(z)$}

\begin{figure}[htb]
\includegraphics[width=0.5\textwidth]{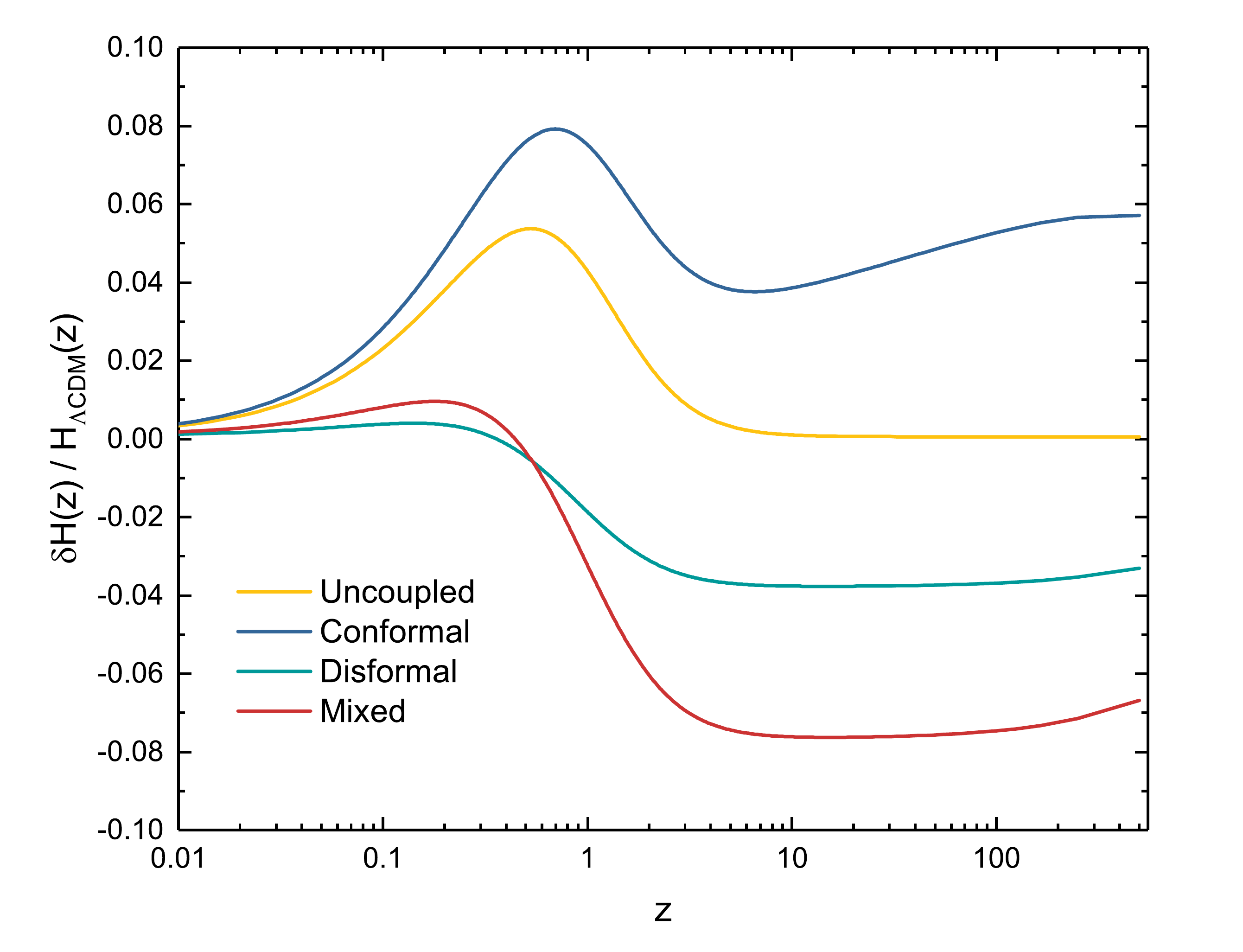}
\caption{\label{fig:delta_H}Deviation of Hubble expansion rate for the coupling model with respect to that of the standard $\Lambda$CDM model, where $\delta H(z)=H(z)-H_{\Lambda \rm CDM}(z)$. Four cases are adopted here to illustrate the evolution of $\delta H/H_{\Lambda \rm CDM}$ as a function of redshift, with the parameter setting described in Tab.~\ref{table:fid}. These curves show that the DM-DE interactions can lead to the relative deviation of $H(z)$ at a few percent level. }
\end{figure}

Referring to Eq.~\ref{eq:t21-1}, the 21 cm brightness temperature $T_{21}(z)$ is a function of three variables $H(z)$, $T_{\rm s}(z)$ and $T_\gamma(z)$. The radiation temperature $T_\gamma(z)$ (assumed to be the same as $T_{\rm CMB}(z)$ in our scenario) evolves in the same way as that in the $\Lambda$CDM model. According to Eq.~\ref{Ts Equation}, the spin temperature $T_{\rm s}$ is tightly coupled with the kinetic temperature $T_{\rm g}$ that further depends on $H(z)$ as well in terms of Eq.~\ref{Tgas Equation}. These dependencies indicate that $H(z)$ is the key factor governing the evolution of $T_{21}$.

In Fig.~\ref{fig:delta_H}, we show the deviation of $H(z)$ in the four fiducial cases described in Tab.~\ref{table:fid}. Compared with $H(z)$ for the standard $\Lambda$CDM model, the coupling model for these four cases can change $H(z)$ at a few percent level on average.  The {\it conformal} and {\it uncoupled} cases effectively increase $H(z)$ over all redshift range and the {\it disformal} and {\it mixed} ones decrease $H(z)$ moderately at early times, especially when $z\gtrsim 1$. These four couplings are rapidly switched on and begin to affect the Hubble parameter at $z\gtrsim 0.1$. The coupling effects become almost negligible at the present epoch, leading to the deviations approach to zero (e.g., less than 1\% level when $z<0.05$), and satisfying our initial conditions to match observed cosmological parameters at present day.

\subsection{Impacts of DM-DE Coupling on $T_{\rm g}$ and $T_{\rm s}$}

\begin{figure}[htb]
\includegraphics[width=0.5\textwidth]{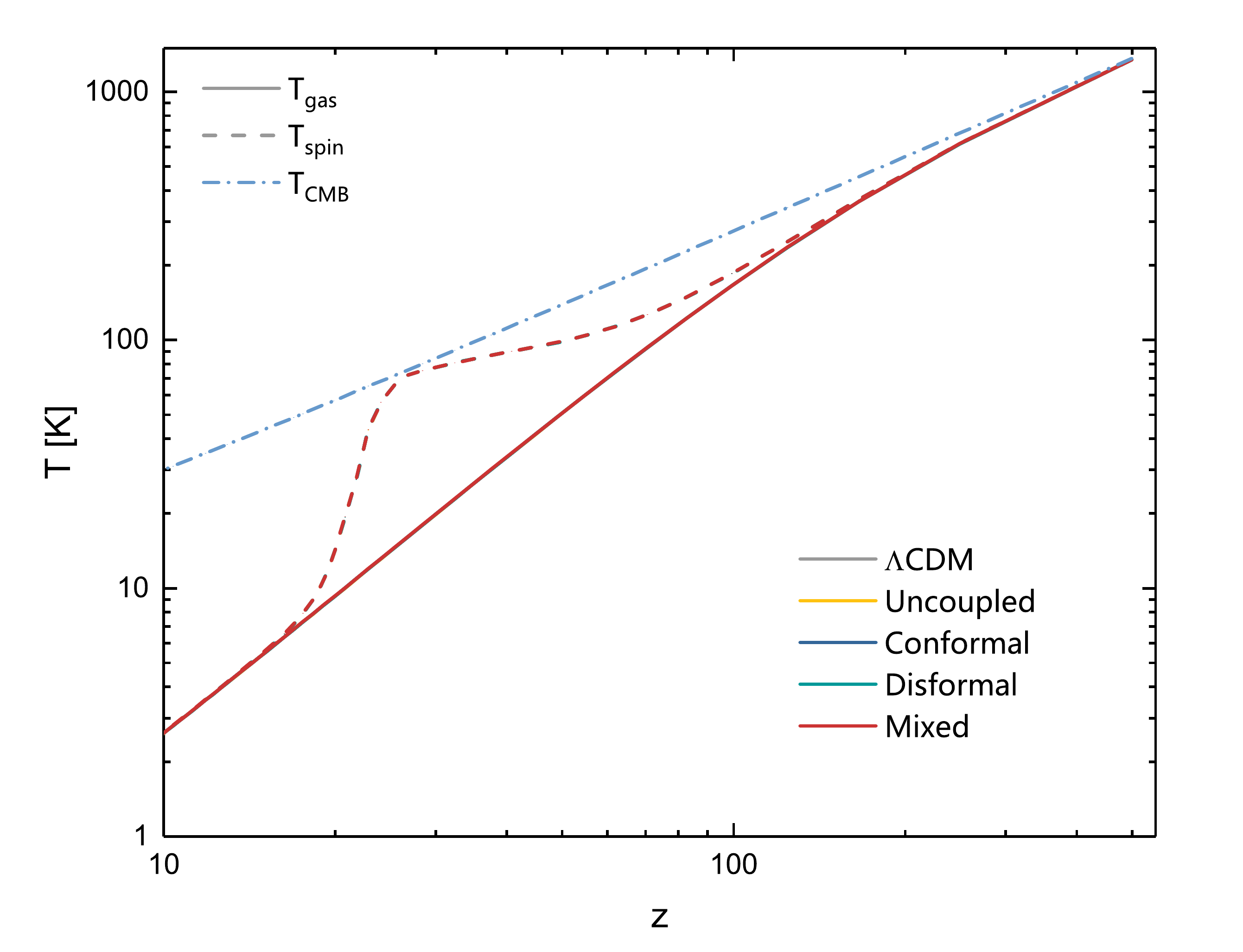}
\caption{\label{fig:TK}Evolution of the CMB temperature $T_{\rm CMB}$ (dash-dotted), the gas kinetic temperature $T_{\rm g}$ (solid), and the spin temperature $T_{\rm s}$ (dash), for the coupling model in fiducial cases of Tab.~\ref{table:fid}. The effects of DE-DM coupling in $T_{\rm g}$ and $T_{\rm s}$ are so small that corresponding curves are overlapped with that from the standard $\Lambda$CDM model. These coupling effects thus can be essentially ignored.
}
\end{figure}

\begin{figure}[htb]
\includegraphics[width=0.5\textwidth]{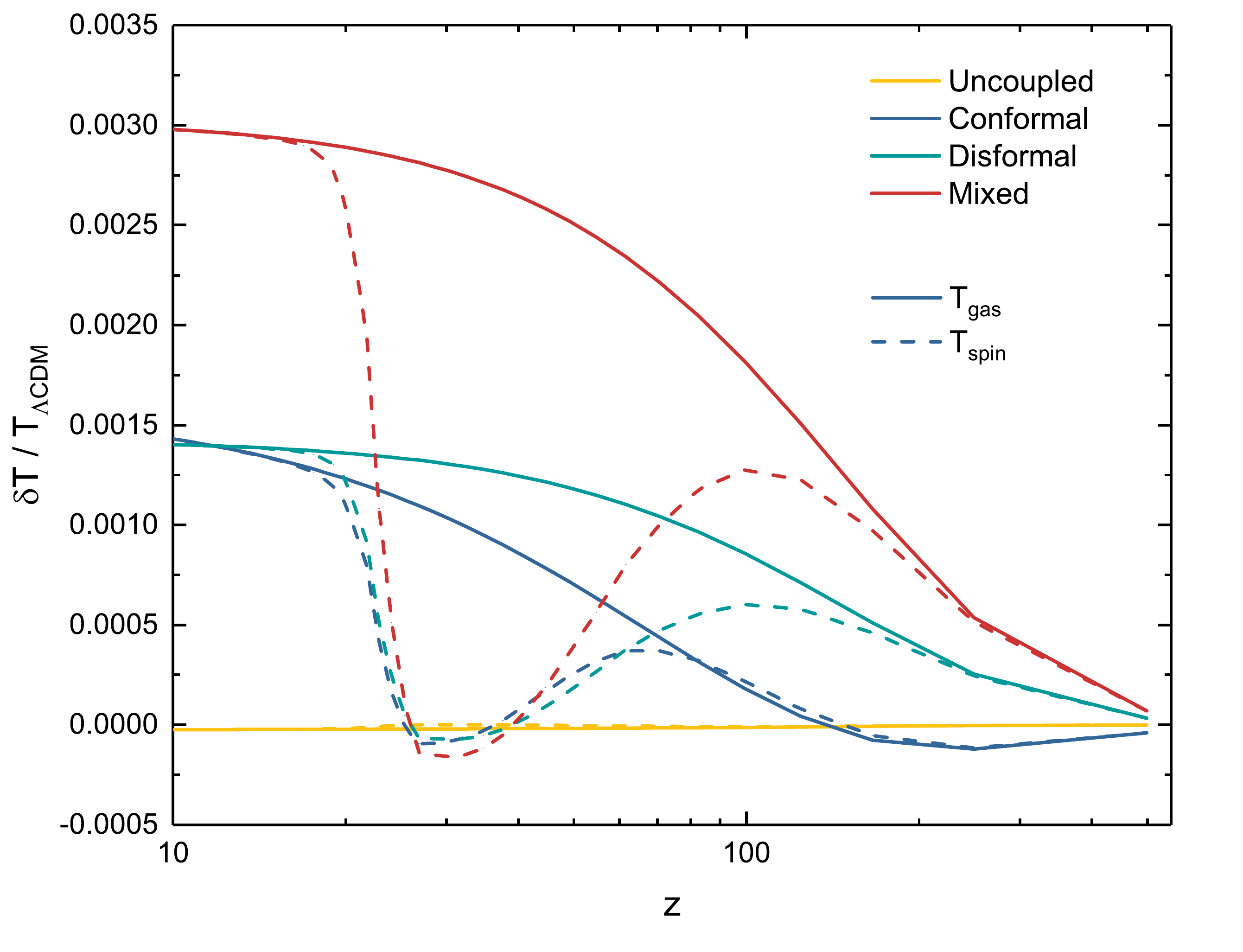}
\caption{\label{fig:delta_T}Same as Fig.~\ref{fig:TK}, but for relative deviations of $T_{\rm g}$ (solid) and $T_{\rm s}$ (dash) with respect to those derived in $\Lambda$CDM. The deviations in {\it uncoupled} is negligibly small and the other cases give rise to increases at the level of $<0.3\%$ in both $T_{\rm g}$ and $T_{\rm s}$, which are about an order of magnitude smaller than the changes in $H(z)$ as in Fig.~\ref{fig:delta_H}}
\end{figure}

We will now study the impacts on the evolution of spin temperature $T_{\rm S}$ and gas temperature $T_{\rm g}$, since they may also affect $T_{21}$ based on Eqs.~\ref{T21 Equation} and ~\ref{Ts Equation}. Because of the Wouthuysen-Field effect, $T_{\rm s}$ sharply approaches $T_{\rm g}$ at the regime $z\lesssim25$ where star formation begins and scattering of Ly$\alpha$ photons with hydrogen atoms thus becomes efficient, leading to a strong coupling between $T_{\rm s}$ and $T_{\rm g}$. However, from Fig.~\ref{fig:TK}, we can see that, the changes due to the couplings in dark sectors  are invisible and can be essentially neglected. To clearly demonstrate the relative changes, we plot the fractional difference from the standard $\Lambda$CDM case in Fig.~\ref{fig:delta_T}, i.e. $\delta T/T =  (T-T_{\Lambda \rm CDM})/T_{\Lambda\rm CDM}$. Both $T_{\rm s}$ and $T_{\rm g}$ in the cases of {\it conformal, disformal, mixed} are increased by 0.2\% on average, and the relative deviation in the {\it uncoupled} case is almost zero.

\begin{figure}[htb]
	\includegraphics[width=0.5\textwidth]{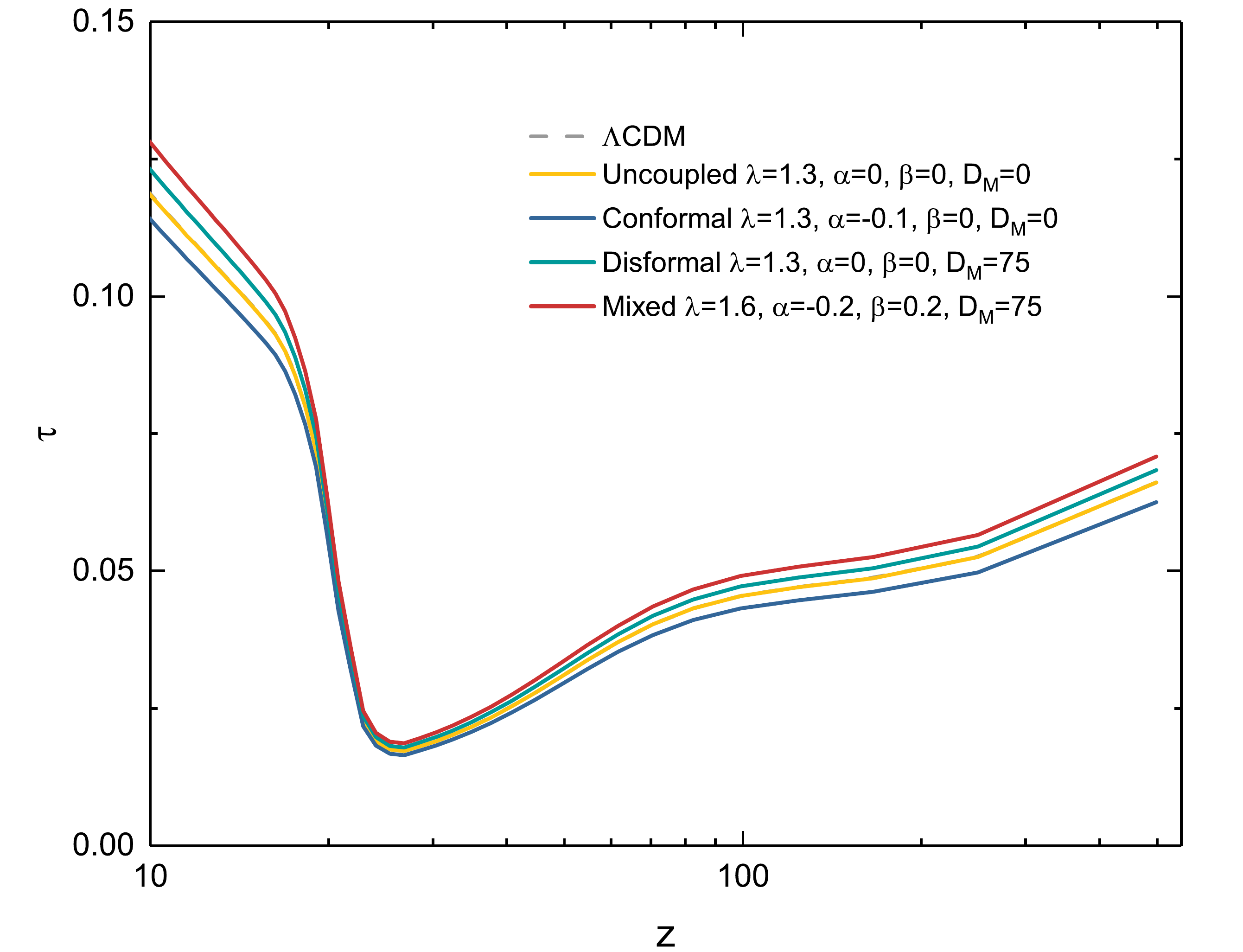}
	\caption{\label{fig:tau} Evolution of the optical depth $\tau$ predicted from the fiducial cases and the $\Lambda$CDM model. There is essentially no difference between the \textit{uncoupled} case (yellow) and the $\Lambda$CDM model (grey dash) as the deviation of $H(z)$ in such case is almost zero at $z\gtrsim10$.}
\end{figure}

In terms of Eq.~\ref{tau Equation}, the optical depth $\tau$ depends on not only $T_{\rm s}$ but also $H(z)$. As mentioned before (see again Figs.~\ref{fig:delta_H} and \ref{fig:delta_T}), since the coupling-induced deviation in $H(z)$ for each case at $z\gtrsim 0.1$ is an order of magnitude greater than that of $T_{\rm s}$, we thus expect $\tau$ would vary by $<10$\% level which should be compatible with the altered amplitude of $H(z)$. In Fig.~\ref{fig:tau}, we present the evolution of $\tau$ for the fiducial cases which confirms our expectation. Except for the {\it uncoupled} case, the other three ones lead to up to 10\% deviations, and there is essentially no difference between the {\it uncoupled} one and the $\Lambda$CDM model (two curves overlapped).

We can conclude that, the altered $H(z)$ from our DM-DE coupling model is the dominant factor in changing $T_{21}$ absorption signal as the deviations of $H(z)$ (a few percent level) are about an order of magnitude greater than those of $T_{\rm s}$.

\subsection{Results on 21cm Brightness Temperature}

\begin{figure}[htb]
	\includegraphics[width=0.5\textwidth]{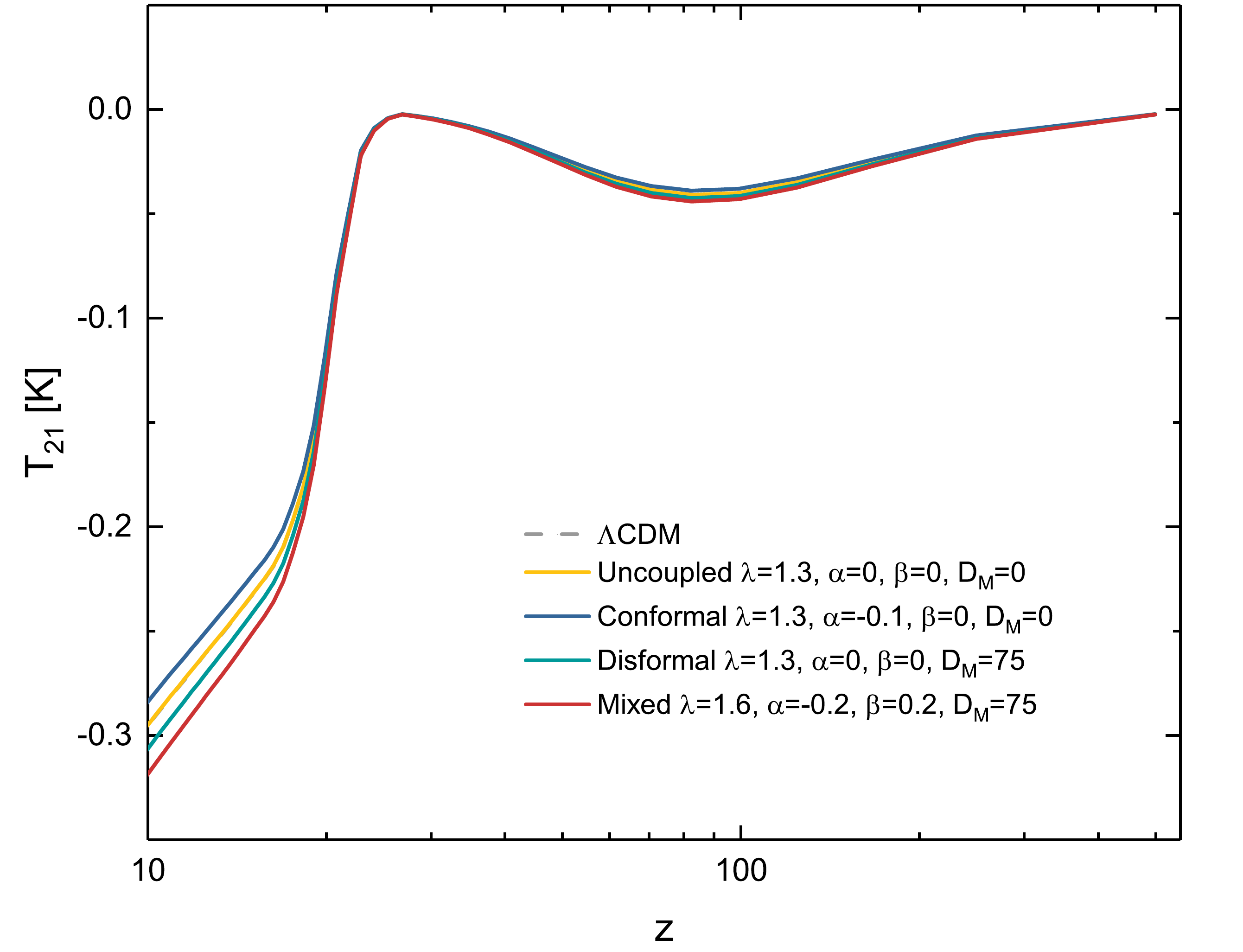}
	\caption{\label{fig:T21}Evolution of 21 cm brightness temperature $T_{21}$ relative to the CMB in the four fiducial coupling cases and the $\Lambda$CDM. The changes of $T_{21}$ at $z\lesssim20$ are mainly due to $H(z)$ altered by the coupling effects. Due to the linear dependence, $T_{21}(z)\propto \tau(z)\sim1/H(z)$, the relative change in $T_{21}$ for each case is the same order of magnitude as that in $\tau(z)$, showing a strong correlation between $T_{21}$ and $\tau$ (see Fig.~\ref{fig:tau}).}
\end{figure}

From the above analysis, we find that the DM-DE coupling in our fiducial cases has more influence on $H(z)$ than on $T_{\rm g}$ and $T_{\rm S}$ at $z\sim 20$, from which we expect this influence may change $T_{21}$ at a comparable level at that epoch. In Fig.~\ref{fig:T21}, we compare the evolution of 21 cm brightness temperature $T_{21}$ in our different coupling cases with that in the $\Lambda$CDM model. As seen, $T_{21}$ at small $z$ ($z\lesssim20$) is indeed altered by about $\pm5\%$ in the {\it conformal, disformal} and {\it mixed} cases, since the changes in $T_{21}$ are closely related to the contribution from $\delta H(z)$ which are about $4\%, -4\%$ and $-8\%$ in relative difference (see Fig.~\ref{fig:delta_H}), respectively.

Next, we turn our attention to investigate whether the DM-DE coupling model can offer the explanation of the EDGES 21 cm anomaly while remaining consistent with other observations. Ref.~\cite{BM2017} has performed a global analysis and placed stringent constraints on the {\it conformal, disformal}, and {\it mixed} DM-DE interactions, by combining current cosmological data from the Planck 2015 observations, BAO measurements and  a supernovae Type Ia sample as well as local values of the Hubble constant, etc. The {\it uncoupled} case is not taken into account here as its impact on $T_{21}$ is negligibly small compared to the other couplings, as mentioned before regarding Fig.~\ref{fig:T21}.

In terms of Eq.~\ref{T21 Equation}, we calculate the predicted values of the 21 cm absorption signal from various coupling cases spanning over a wide range of parameters listed in Tab.~\ref{table:parameters}.  We have chosen the usual parameter convention described in Ref.~\cite{BM2017} to derive the signals from {\it conformal, disformal}, and {\it mixed} cases. In our calculations, we focus on the peak redshift of the absorption trough in the EDGES result, and the redshift is thus fixed to be $z=17.5$.

\begin{table}
	\begin{center}
		\begin{tabular}{ c  c  c}
			\hline
			\hline
			Parameter  & Space~\\
			\hline
			$\lambda$\dotfill & $[0,\,2.3]$\\
			$\alpha$\dotfill & $[-0.96,\,0.96]$ \\
			$\beta$\dotfill & $[-6,\,3]$ \\
			$D_m/{\rm Mpc}^{1/2}$\ldots & $[0,\,330]$ \\
			\hline
			\hline
		\end{tabular}
	\end{center}
	\caption{\label{table:parameters} The parameter space of the three different coupling cases adopted for analysis.  Taking values outside of the range of parameters is not allowed by the initial conditions described in Sect.~\ref{sec:results}.}
\end{table}

\begin{figure}[htb]
	\includegraphics[width=0.5\textwidth]{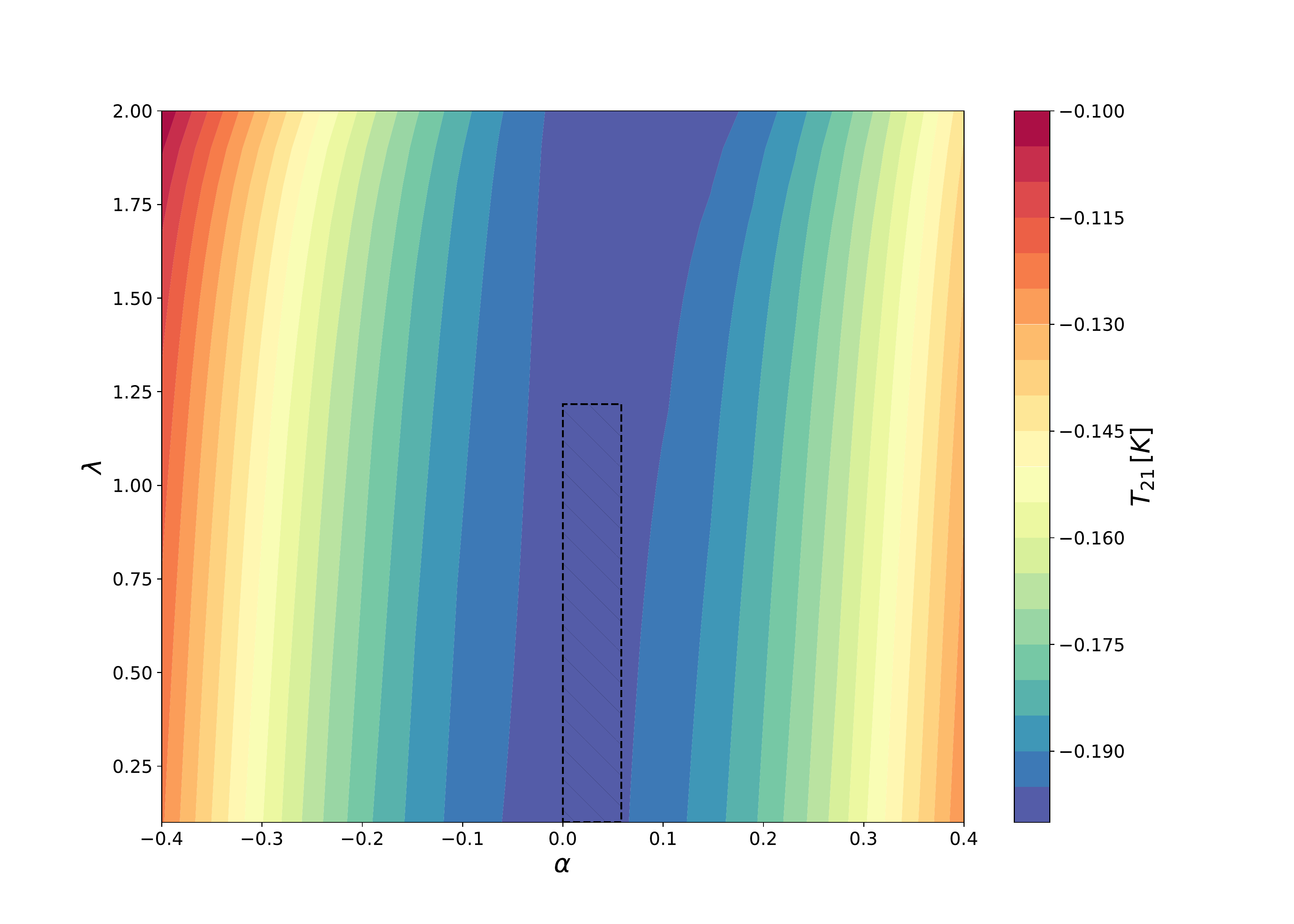}
	\caption{\label{fig:Conformal} 21 cm brightness temperature $T_{21}$ as the function of the interaction parameters in the {\it conformal} case at $z=17.5$. The hatched region corresponds to the 1-$\sigma$ allowed parameter space obtained by a joint analysis from cosmological observations~\cite{BM2017} (same for Fig.~\ref{fig:Disformal} \&~\ref{fig:Mixed}), where only the positive priors on the coupling parameters (listed in Tab.~\ref{table:parameters}) are considered. Parameters not satisfying the initial conditions as mentioned before are excluded in our analysis. The resulting signals are much weaker than the expected one ($\simeq-0.5$K), with the minimum value of $T_{21}=-0.2$K.}
\end{figure}

In the {\it conformal} case where $\beta$ and $D_m $ are set to be zero by definition~\cite{BM2017}, we systematically explore the parameter space of $\alpha$ and $\lambda$. The resulting values of $T_{21}$ are shown in Fig.~\ref{fig:Conformal}, where the hatched area corresponds to the 1-$\sigma$ allowed region of ~\cite{BM2017} inferred from a joint analysis of cosmological observations. As seen,  the {\it conformal} coupling can not suppress $T_{21}$ down to the EDGES-observed amplitude ($\simeq-0.5$K) which is still about 2.5 times smaller than the minimum value of $T_{21}=-0.2$K derived from such coupling. The derived 21 cm signal in the hatched region is almost the same as the typical value predicted from the $\Lambda$CDM model.

\begin{figure}[htb]
	\includegraphics[width=0.5\textwidth]{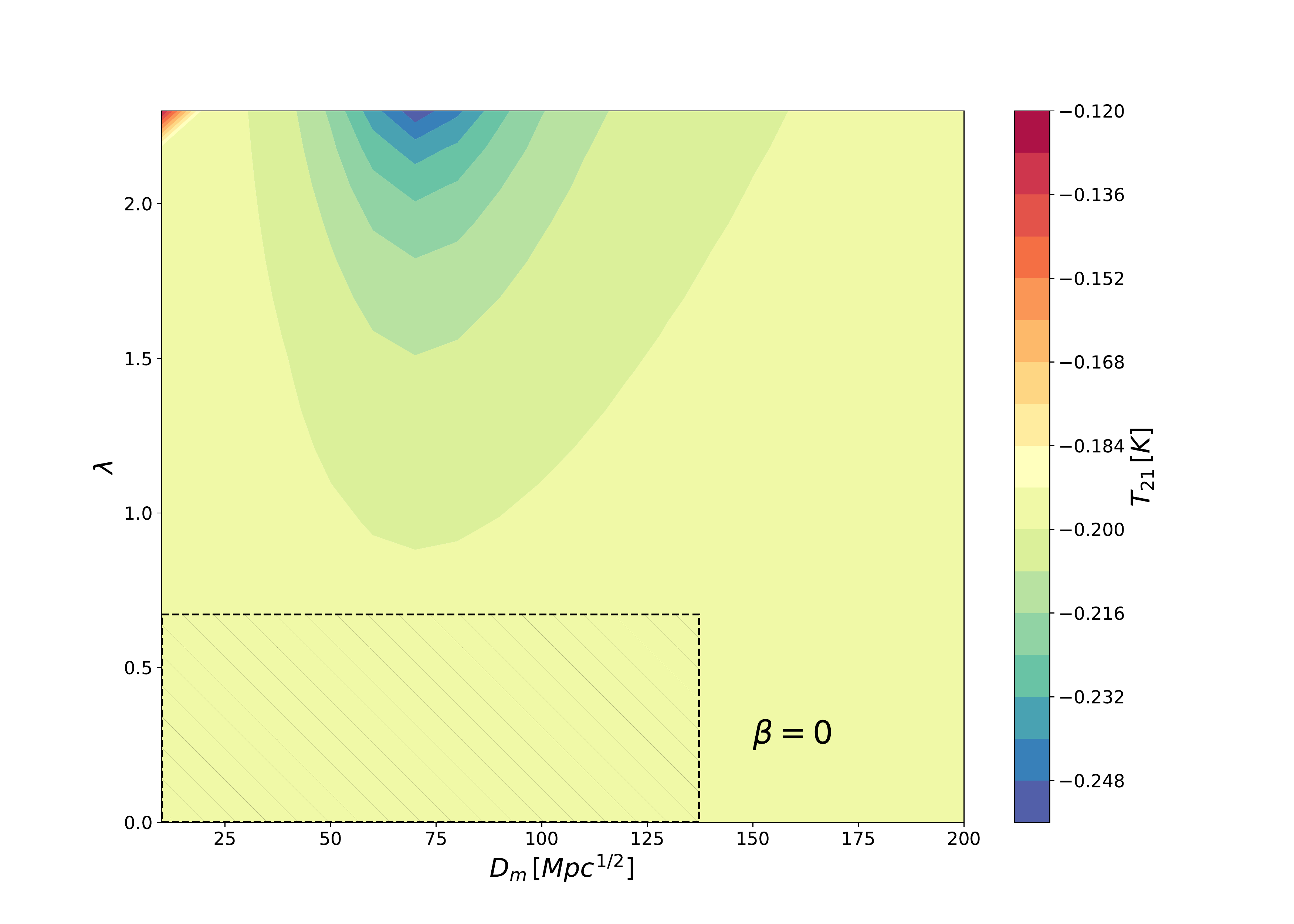}\\
	\includegraphics[width=0.5\textwidth]{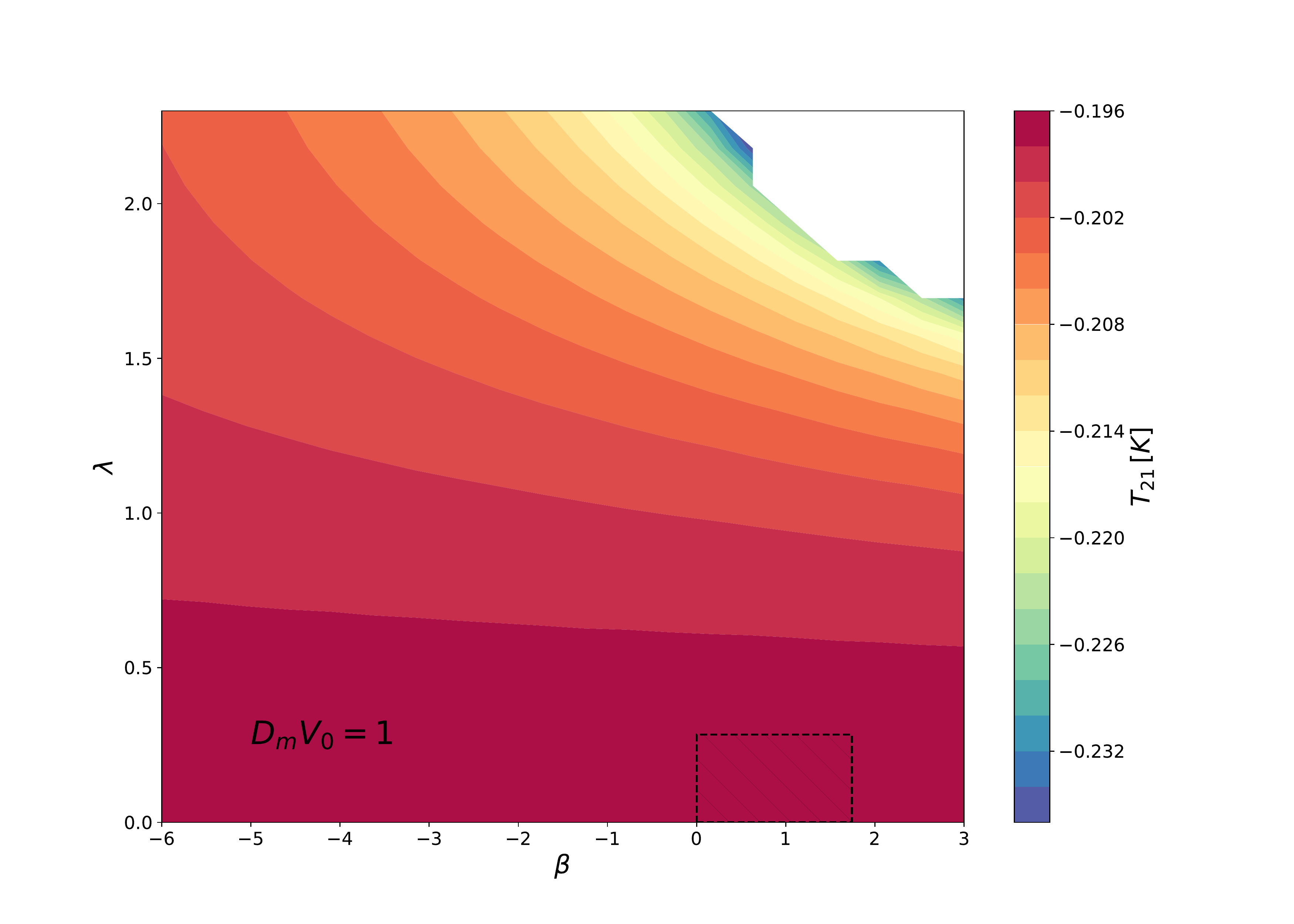}
	\caption{\label{fig:Disformal} Same as Fig.~\ref{fig:Conformal}, but for the {\it disformal} case where $\beta$ and $D_mV_0$ are set to be $\beta=0$ (upper panel) and $D_mV_0=1$ (lower panel). The parameters within the blank region are excluded by the constraints from our initial conditions. The minimum value of $T_{21}=-0.25$ K is achieved when $\lambda=2.3, D_m=70~\rm{Mpc}^{1/2}$ (upper), and of $T_{21}= -0.24$K when $\lambda=2.2, \beta=0.6$ (lower).}
\end{figure}

\begin{figure*}[htpb]
\centering
\mbox{
 \subfigure[$\lambda=2$]{
   \includegraphics[width=3.3in] {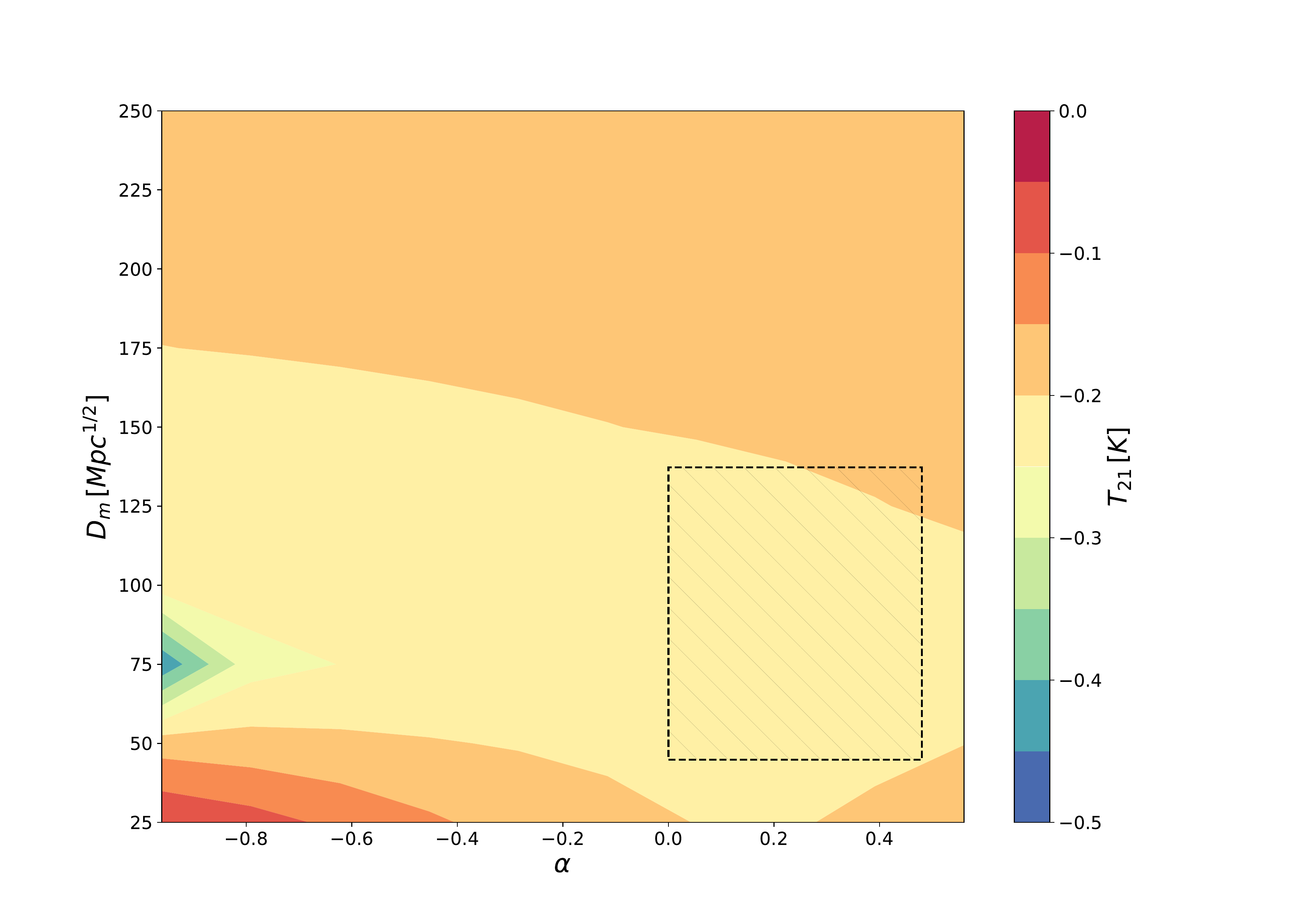}
    }

\subfigure[$D_m=75~\rm{Mpc}^{1/2}$]{
   \includegraphics[width=3.3in] {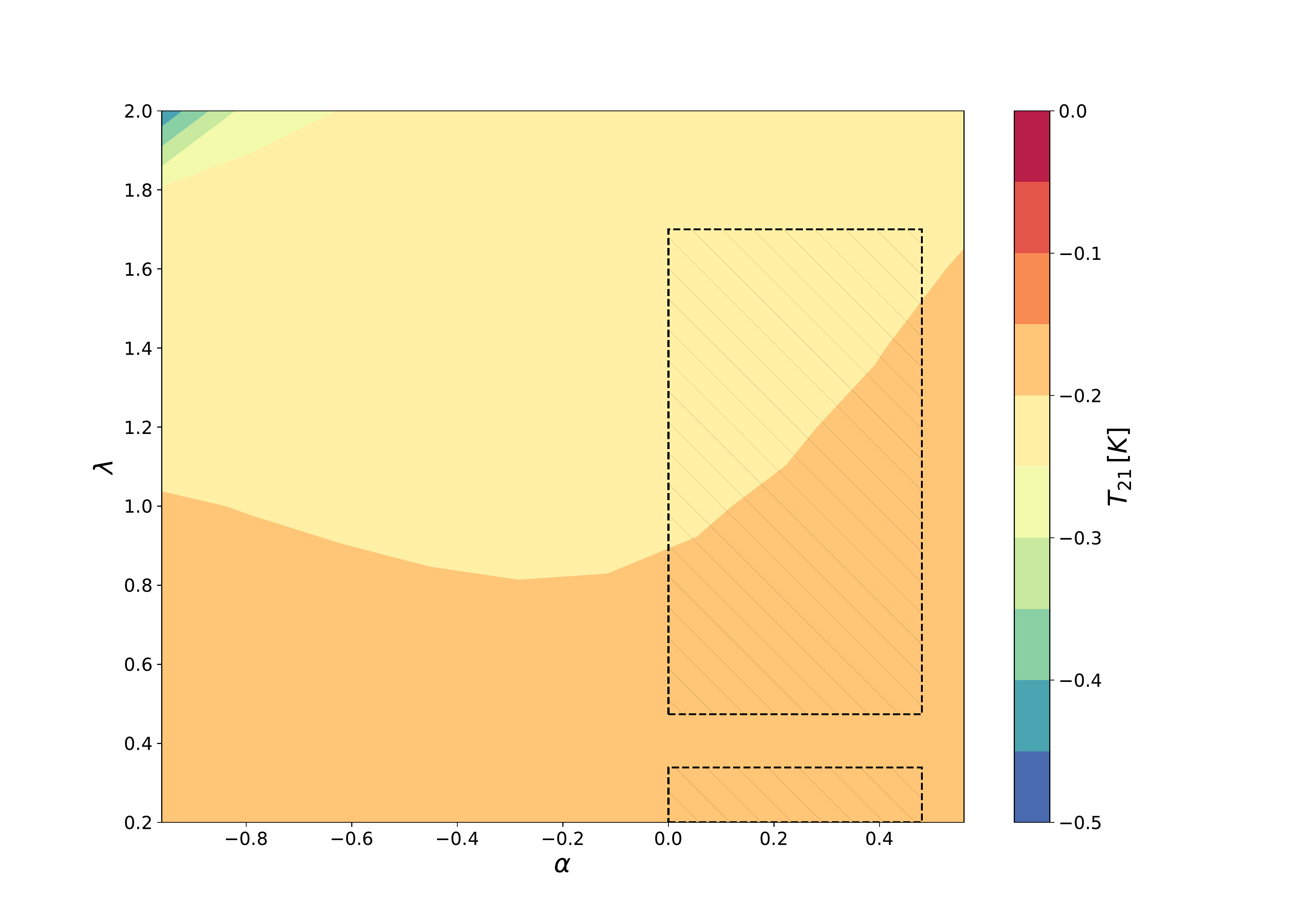}
}
}
\mbox{
\subfigure[$\alpha=-0.96$]{
   \includegraphics[width=3.3in] {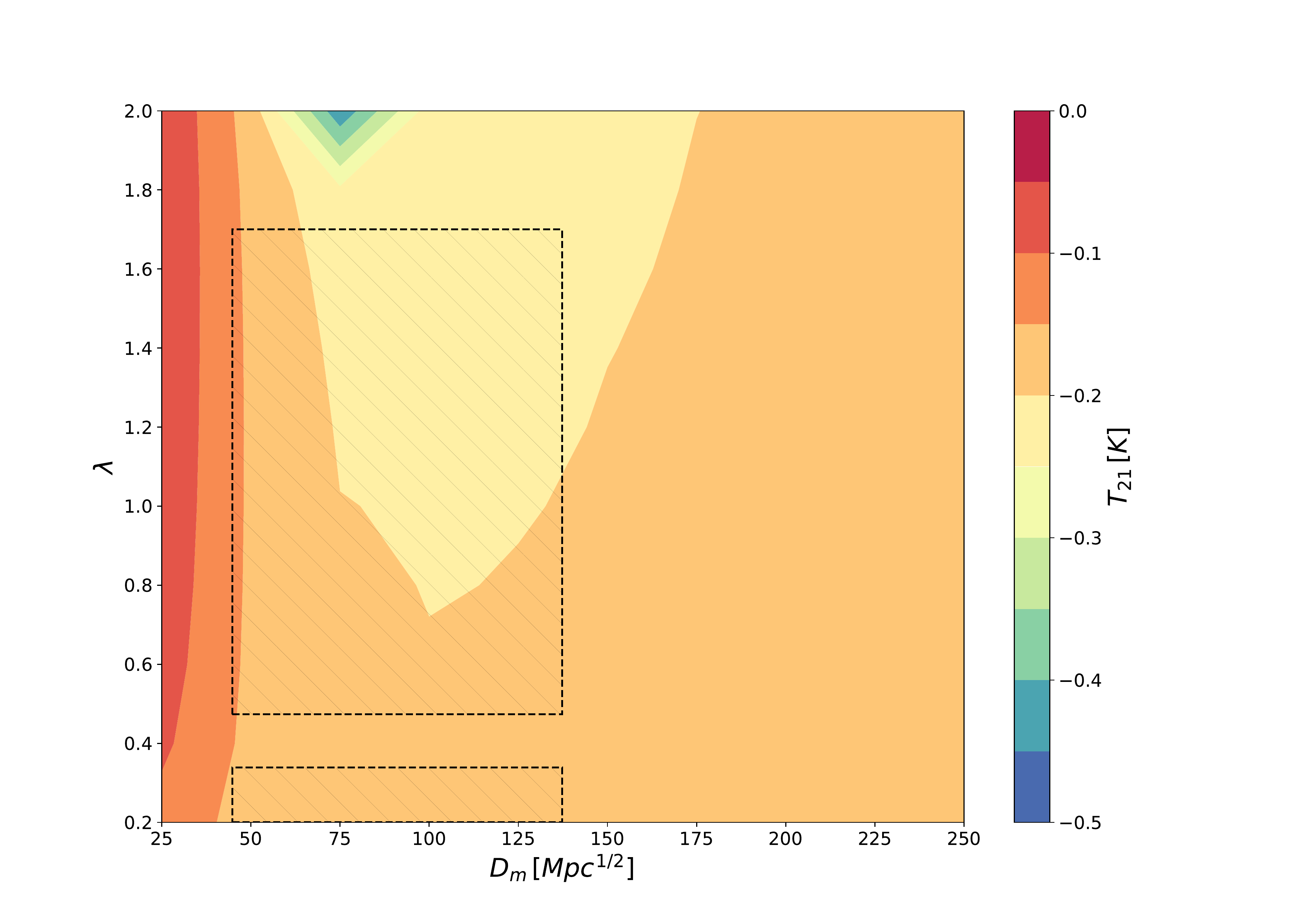}
}
}

\caption{\label{fig:Mixed} Same as Fig~\ref{fig:Conformal}, but for the {\it mixed} case, which has a constant {\it disformal} coupling by setting $\beta=0$~\cite{BM2017}. The mixed coupling can suppress $T_{21}$ down to $-0.44$ K when $\lambda=2, \alpha=-0.96$ and $D_m=75~\rm{Mpc}^{1/2}$, approaching to the EDGES measurement, whereas the values of the parameters favored by EDGES will lead to $\rho_c$ at $z\simeq1100$ inconsistent with the current cosmological measurements at least at 3-$\sigma$.}
\end{figure*}

Note that, the obtained cosmological constraints~\cite{BM2017} (corresponding to the hatched regions in Figs.~\ref{fig:Conformal},~\ref{fig:Disformal},~\ref{fig:Mixed}) on the DM-DE interaction model are based a positive prior for the coupling parameters $\alpha$ and $\beta$. In this study, we allow these two parameters to vary freely since there are no any physical reasons to restrict the negative values.

In Fig.~\ref{fig:Disformal}, we present the evolution of $T_{21}$ at $z=17.5$ for the {\it disformal} case by varying the values of $\lambda$ and $D_m$ (upper panel), and, $\lambda$ and $\beta$ (lower panel), respectively, corresponding a ``constant-coupling'' interaction with $\beta = 0$ and $\alpha = 0$,  and a ``time-dependent coupling'' one with $D_mV_{0}=1$ and $\alpha = 0$. The predicted largest absorption signal of $T_{21}=-0.25$ K are still about two times smaller than the detected one by EDGES. Furthermore, the predicted $T_{21}$  ($\approx -0.24$ K) near the blank region in the lower panel of Fig.~\ref{fig:Disformal} tends to become more negative, close to the expected value, indicating that the {\it disformal} coupling may allow one to reproduce EDGES 21 cm anomaly if there is no restriction on the initial conditions (i.e., if relaxing $\Omega_{i}$). In addition, taking into account the parameters in the hatched region to match the cosmological measurements within 1-$\sigma$, one can obtain a smaller absorption signal where  $T_{21} \approx -0.196$ K, which is much disfavored by EDGES.

In Fig.~\ref{fig:Mixed}, we illustrate the derived $T_{21}$ signals in the {\it mixed} coupling case, by varying the free parameters $D_m, \alpha$ and $\lambda$, while with $\beta=0$ fixed, corresponding to a constant {\it disformal} coupling (the same parameter choice as in ~\cite{BM2017}). For each plot in Fig.~\ref{fig:Mixed}, we vary only two of them and keep the other one fixed such that they can clearly show the dependence of the parameters on $T_{21}$.

As seen, the largest absorption signal at $z=17.5$ can reach about $-0.44$ K, which is essentially matched with the EDGES result. Compared to the {\it conformal} and {\it disformal} coupling cases, the {\it mixed} one leads to relatively larger absorption signal, which is expected since the {\it mixed} coupling decreases $H(z)$ more efficiently than the other cases and thus yields such more negative 21 cm signal (see Figs.~\ref{fig:delta_H} and~\ref{fig:T21}). However, the EDGES favored range for the coupling parameters is still in conflict with present cosmological data. We find that the parameters in the {\it mixed} case that produce the largest 21cm absorption signal will significantly change $\rho_c$ at $z\simeq1100$, ruled out by the current CMB measurements at least at 3-$\sigma$ level. The predicted $T_{21}$ ($\sim0.2$ K) in the hatched region allowed by~\cite{BM2017} remains well above the expected one, comparable with the predictions from the {\it conformal} and {\it disformal} cases.

Based on the above analysis, we find that only the {\it mixed} coupling case can in principle produce an amplitude of 21 cm absorption signal comparable with the EDGES measurement. However, the values of the coupling parameters that are necessary to explain the EDGES anomaly tends to generate a serious conflict with the constraints from present observations. In short, the interacting DM-DE model considered here would not be able to explain the EDGES anomaly while being consistent with present cosmological probes.

\section{Conclusions}
\label{sec:conclusions}

The recent measurement of an excess in the 21-cm brightness temperature from cosmic dawn by the EDGES team has attracted a wide attention, as the observed signal is much deeper than expected in the standard $\Lambda$CDM universe. Various mechanisms have been proposed to alleviate such tension.

In this study, we investigate the possibility of using a generic interacting DM-DE model to explain the EDGES anomaly by calculating the impact of such coupling model on the 21 cm absorption signal.

As known, the big-bang nucleosynthesis (BBN) provides a stringent constraint on any nonstandard cosmology scenarios. As the predicted abundance of light elements from the standard BBN is in well agreement with observations, any deviations of the baryon-to-photon ratio  and the Hubble rate $H(z)$ during the BBN epoch would significantly change the the abundance of the light elements. Since the proposed  DM-DE interaction model keeps the standard behavior of baryons unchanged from the early universe before the onset of BBN ($z\leq10^9$) and only introduces an extra coupling between DM and DE, the influence from such interactions on the baryon-to-photon ratio is hence negligible small. Furthermore, we also find that the deviations of $H(z)$ induced by those nonstandard coupling terms are at the level of $\delta H(z)/H_{\rm \Lambda CDM}(z)<10^{-4}$ during the BBN epoch ($z\sim 10^8$). Therefore, the DM-DE coupling in our study gives only negligible small effects in the nucleosynthesis process and hence we can safely neglect the constraints from the BBN. However, one should be cautious when dealing with any nonstandard terms that could change the baryon-to-photon ratio and the Hubble rate significantly during BBN epoch.

We find the interaction with the fiducial cases can effectively change the Hubble expansion rate $H(z)$ at a few percent level at redshifts of $10\lesssim z \lesssim20$, but the changes in the 21 cm spin temperature $T_{\rm s}$ are about an order of magnitude smaller than those in $H(z)$, which implies that the coupling-induced effects on $T_{21}$ are mostly contributed from the variations of $H(z)$ rather than $T_{\rm s}$.

Finally, we explore a wide range of the parameter space for the various couplings, including {\it disformal, conformal, uncoupled} and {\it mixed} ones, to further investigate whether this coupling model can offer an explanation of the EDGES 21 cm anomaly while being consistent with the limits placed by a joint analysis from present cosmological observations. We find that, the derived $T_{21}$ from the {\it disformal, conformal} and {\it uncoupled} couplings are much weaker than  the EDGES results. Only the {\it mixed} case can lead to an amplitude with the maximum value of $-0.44$ K at $z=17.5$, almost consistent with the expected value of about $-0.5$ K from EDGES. Unfortunately, the region of the {\it mixed} coupling parameters allowed by EDGES is in conflict with the other cosmological observations. Definitely, 21 cm cosmology offers the exciting opportunities to probe/test DM-DE interactions and future 21 cm observations at cosmic dawn would bring us more insight to understand the non-standard cosmological effects.

\vspace{1.0em}
\section*{Acknowledgments}
This work was supported by the National Science Foundation of China (11621303, 11653003,11773021), and the National Key R\&D Program of China (2018YFA0404601). BY also acknowledges the support of  the Hundred Talents (Young Talents) program from the CAS and the NSFC-CAS  joint fund for space scientific satellites No. U1738125. YX acknowledges the support by the Chinese Academy of Sciences (CAS) Strategic Priority Research Program XDA15020200, the CAS Frontier Science Key Project QYZDJ-SSW-SLH017, the National Natural Science Foundation of China (NSFC) key project grant 11633004, and the NSFC-ISF joint research program No. 11761141012. B W's work was partially supported by NNSFC. 



\begin{thebibliography}{99}
\bibitem[Bowman et al.(2018)]{EDGES} Bowman, J.~D., Rogers, A.~E.~E., Monsalve, R.~A., Mozdzen, T.~J., \& Mahesh, N.\ 2018, \nat, 555, 67
\bibitem[Barkana(2018)]{Barkana2018Nature} Barkana, R.\ 2018, \nat, 555, 71
\bibitem[Wouthuysen(1952)]{Wouthuysen1952} Wouthuysen, S.~A.\ 1952, \aj, 57, 31
\bibitem[Field(1959)]{Field1959} Field, G.~B.\ 1959, \apj, 129, 536
\bibitem[Cohen et al.(2017)]{Cohen2017} Cohen, A., Fialkov, A., Barkana, R., \& Lotem, M.\ 2017, \mnras, 472, 1915
\bibitem[Mu{\~n}oz \& Loeb(2018)]{ML2018} Mu{\~n}oz, J.~B., \& Loeb, A.\ 2018, arXiv:1802.10094
\bibitem[Fialkov et al.(2018)]{FBC2018} Fialkov, A., Barkana, R., \& Cohen, A.\ 2018, arXiv:1802.10577
\bibitem[Berlin et al.(2018)]{BHKM2018} Berlin, A., Hooper, D., Krnjaic, G., \& McDermott, S.~D.\ 2018, arXiv:1803.02804
\bibitem[Barkana et al.(2018)]{BORV2018} Barkana, R., Outmezguine, N.~J., Redigolo, D., \& Volansky, T.\ 2018, arXiv:1803.03091
\bibitem[Fraser et al.(2018)]{FHH2018} Fraser, S., Hektor, A., H{\"u}tsi, G., et al.\ 2018, arXiv:1803.03245
\bibitem[Li \& Cai(2018)]{LC2018} Li, C., \& Cai, Y.-F.\ 2018, arXiv:1804.04816
\bibitem[Cheung et al.(2018)]{CKNT2018} Cheung, K., Kuo, Jui-Lin., Ng, Kin-Wang., \& Tsai, Yue-lin, S.\ 2018 arXiv:1803.09398
\bibitem[Clark et al.(2018)]{CDGMS2018} Clark, S., Dutta, B., Gao, Y., Ma, Y.-Z., \& Strigari, L.~E.\ 2018, arXiv:1803.09390
\bibitem[Ewall-Wice et al.(2018)]{ECL2018} Ewall-Wice, A., Chang, T.-C., Lazio, J., et al.\ 2018, arXiv:1803.01815
\bibitem[Costa et al.(2018)]{Costa2018} Costa, A.~A., Landim, R.~C.~G., Wang, B., \& Abdalla, E.\ 2018, arXiv:1803.06944
\bibitem[Hill \& Baxter(2018)]{HB2018} Hill, J.~C., \& Baxter, E.~J.\ 2018, arXiv:1803.07555
\bibitem[Wang et al.(2016)]{WAAP2016} Wang, B., Abdalla, E., Atrio-Barandela, F., \& Pav{\'o}n, D.\ 2016, Reports on Progress in Physics, 79, 096901
\bibitem[Karwal \& Kamionkowski(2016)]{KK2016} Karwal, T., \& Kamionkowski, M.\ 2016, \prd, 94, 103523
\bibitem[Planck Collaboration et al.(2016)]{Planck2015-13} Planck Collaboration, Ade, P.~A.~R., Aghanim, N., et al.\ 2016, \aap, 594, A13
\bibitem[Wang \& Zhao(2018)]{WZ2018} Wang, Y., \& Zhao, G.-B.\ 2018, arXiv:1805.11210
\bibitem[Husain \& Qureshi(2016)]{HQ2016} Husain, V., \& Qureshi, B.\ 2016, Physical Review Letters, 116, 061302
\bibitem[van de Bruck \& Morrice(2015)]{BM2015} van de Bruck, C., \& Morrice, J.\ 2015, \jcap, 4, 036
\bibitem[Mifsud \& van de Bruck(2017)]{MB2017} Mifsud, J., \& van de Bruck, C.\ 2017, \jcap, 11, 001
\bibitem[van de Bruck \& Mifsud(2018)]{BM2017} van de Bruck, C., \& Mifsud, J.\ 2018, \prd, 97, 023506
\bibitem[Koivisto et al.(2012)]{KMZ2012} Koivisto, T.~S., Mota, D.~F., \& Zumalac{\'a}rregui, M.\ 2012, Physical Review Letters, 109, 241102
\bibitem[Zumalac{\'a}rregui et al.(2013)]{ZKM2012} Zumalac{\'a}rregui, M., Koivisto, T.~S., \& Mota, D.~F.\ 2013, \prd, 87, 083010
\bibitem[D'Amico et al.(2016)]{DAHK2016} D'Amico, G., Hamill, T., \& Kaloper, N.\ 2016, \prd, 94, 103526
\bibitem[Lewis et al.(2000)]{LCL2000} Lewis, A., Challinor, A., \& Lasenby, A.\ 2000, \apj, 538, 473
\bibitem[Seager et al.(1999)]{Seager1999} Seager, S., Sasselov, D.~D., \& Scott, D.\ 1999, \apjl, 523, L1
\bibitem[Zaldarriaga et al.(2004)]{Zaldarriaga2004} Zaldarriaga, M., Furlanetto, S.~R., \& Hernquist, L.\ 2004, \apj, 608, 622
\bibitem[Xu et al.(2018)]{XYC2018} Xu, Y., Yue, B., \& Chen, X.\ 2018, arXiv:1806.06080
\bibitem[Ciardi \& Madau(2003)]{CM2003} Ciardi, B., \& Madau, P.\ 2003, \apj, 596, 1
\bibitem[Furlanetto et al.(2006)]{FOB2006} Furlanetto, S.~R., Oh, S.~P., \& Briggs, F.~H.\ 2006, \physrep, 433, 181
\bibitem[Chluba \& Thomas(2011)]{CT2011} Chluba, J., \& Thomas, R.~M.\ 2011, \mnras, 412, 748
\bibitem[Rubi{\~n}o-Mart{\'{\i}}n et al.(2010)]{Rubino-Martin2010} Rubi{\~n}o-Mart{\'{\i}}n, J.~A., Chluba, J., Fendt, W.~A., \& Wandelt, B.~D.\ 2010, \mnras, 403, 439
\bibitem[Chluba(2010)]{Chluba2010} Chluba, J.\ 2010, \mnras, 402, 1195
\bibitem[Chluba et al.(2010)]{CVD2010} Chluba, J., Vasil, G.~M., \& Dursi, L.~J.\ 2010, \mnras, 407, 599
\bibitem[Costa et al.(2015)]{COE2015} Costa, A.~A., Olivari, L.~C., \& Abdalla, E.\ 2015, \prd, 92, 103501




\end{thebibliography}
\end{document}